\documentclass[amsmath, amssymb, preprintnumbers, showpacs, showkeys, aps,prb,superscriptaddress,twocolumn]{revtex4-1}

\pdfoutput=1 
\usepackage{amssymb}
\usepackage{comment}
\usepackage[utf8]{inputenc}
\usepackage{graphicx}
\usepackage{amsmath}
\usepackage{csquotes}
\usepackage{xcolor}
\usepackage{placeins}
\usepackage[pdftex,colorlinks=false]{hyperref}
\hypersetup{
    colorlinks,
    citecolor=black,
    filecolor=black,
    linkcolor=black,
    urlcolor=black
}
\usepackage{tikz}

\definecolor{lime}{HTML}{A6CE39}
\DeclareRobustCommand{\orcidicon}{
	\begin{tikzpicture}
	\draw[lime, fill=lime] (0,0) 
	circle [radius=0.16] 
	node[white] {{\fontfamily{qag}\selectfont \tiny ID}};
	\draw[white, fill=white] (-0.0625,0.095) 
	circle [radius=0.007];
	\end{tikzpicture}
	\hspace{-2mm}
}
\foreach \x in {A, ..., Z}{\expandafter\xdef\csname orcid\x\endcsname{\noexpand\href{https://orcid.org/\csname orcidauthor\x\endcsname}
			{\noexpand\orcidicon}}
}

\usepackage[globalcitecopy]{bibunits}
\defaultbibliography{bibliography}
\defaultbibliographystyle{apsrev4-1}

\def\bs#1{\boldsymbol{#1}}	    
\def\imi{\mathrm{i}}		    

\DeclareMathAlphabet{\mathbbold}{U}{bbold}{m}{n}

\begin{document}

\title{Analysis of Charge Order in the Kagome Metal $A$V$_3$Sb$_5$ ($A=$K,Rb,Cs)}

\author{M.~Michael~Denner\orcidA{}}
\affiliation{Department of Physics, University of Zurich, Winterthurerstrasse 190, 8057 Zurich, Switzerland}
\author{Ronny Thomale\orcidB{}}
\affiliation{Institute for Theoretical Physics, University of Würzburg, Am Hubland, D-97074 Würzburg, Germany}
\affiliation{Department of Physics and Quantum Centers in Diamond and Emerging Materials (QuCenDiEM) group, Indian Institute of Technology Madras, Chennai 600036, India}
\author{Titus Neupert\orcidC{}}
\affiliation{Department of Physics, University of Zurich, Winterthurerstrasse 190, 8057 Zurich, Switzerland}

\begin{abstract}
Motivated by the recent discovery of unconventional charge order, we
develop a theory of electronically mediated charge density wave formation in the family of kagome
metals $A$V$_3$Sb$_5$ ($A=$K,Rb,Cs). The intertwining of van Hove filling and sublattice
interference suggests a three-fold charge density wave instability at
T$_{\text{CDW}}$. From there, the
charge order forming below T$_{\text{CDW}}$  
can unfold into a variety of phases capable of exhibiting orbital
currents and nematicity. We develop a
Ginzburg Landau formalism to stake out the parameter space of kagome
charge order. We find a nematic chiral charge order to be
energetically preferred, which shows tentative agreement with
experimental evidence. 
\end{abstract}
\maketitle

{\it Introduction.} Density wave instabilities describe the onset of 
translation symmetry breaking of the charge or spin distribution in a
Fermi liquid. While spin density waves usually unambiguously derive from electronic
interactions, it is often subtle to tell whether electronic charge
order is cause or consequence~\cite{RevModPhys.60.1129,cdw}. For the latter, structural transitions
of any kind can accordingly manifest themselves in the
rearranged electronic charge profile. For the former, the electron
fluid minimizes its energy by translation symmetry breaking, which is
mediated by electronic interactions. With regard to the charger order
in the kagome metal $A$V$_3$Sb$_5$ ($A = $K, Rb, Cs)~\cite{PhysRevMaterials.3.094407,jiang2020discovery,arXiv2103.09769,arXiv210309188,arXiv210303118,tan2021charge,uykur2021optical,arXiv210210959,arXiv210210987,arXiv210110193}, a central
observation to begin with is that it coincides with a Fermiology
close to van Hove filling, hinting at a significant enhancement of
electronic correlation effects. While it will still be essential to
further analyse the phonon profile of the material~\cite{tan2021charge,arXiv2103.09769}, we take these observations to
motivate our assumption that the supposed charge order observed in
$A$V$_3$Sb$_5$ is electronically mediated, and that we will constrain
ourselves to the electronic degrees of freedom in the following.

The kagome Hubbard model at van Hove filling has been predicted to
yield a charge density wave (CDW) instability with finite angular
momentum~\cite{PhysRevLett.110.126405,PhysRevB.87.115135}. Generically, a charge density wave interpreted as a
condensate of particle-hole singlets should tend to have zero angular
momentum, in order for the particle hole pair to minimize its energy
with respect to the screened Coulomb interactions. Instead, the van Hove-filling kagome
 Fermi surface yields three nesting vectors, all of which give
rise to individual charge order components. Individually, the particle-hole pair wave function attains angular momentum $l=1$, reminiscent of the
generalization of the Peierls instability to two spatial
dimensions also named charge bond order (CBO). The reason this instability is preferred over other
instabilities roots in the sublattice interference~\cite{PhysRevB.86.121105}, which effectively
increases the relevance of nearest neighbor over on-site Coulomb
repulsion within the nesting channels. 
\begin{figure}[ht!]
    \includegraphics[width=\linewidth]{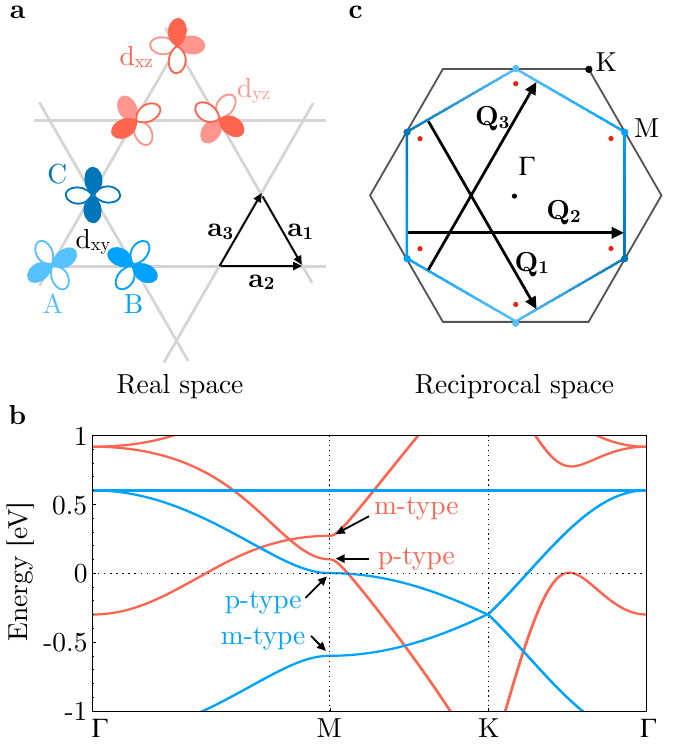}
    \caption{\label{fig:lat-bndstruc} \textsf{\textbf{Electronic features of AV$_3$Sb$_5$.} (\textbf{a}) Kagome lattice with basis of three sublattices, occupied by vanadium $d_{\mathrm{xy}}$ (blue) and $d_{\mathrm{xz/yz}}$ (red) orbitals. (\textbf{b}) Corresponding band structure induced by $d_{\mathrm{xy}}$ (blue) and $d_{\mathrm{xz/yz}}$ (red) orbitals and corresponding nature of van Hove singularities. Higher-lying bands of $d_{\mathrm{xz/yz}}$ are not displayed for simplicity. (\textbf{c}) Schematic of Fermi surfaces in the hexagonal Brillouin zone. The V $d_{\mathrm{xy}}$ Fermi surface, which is nested by the ordering wave vectors $\bs{Q}_j$, $j=1,2,3$, has weight on distinct sublattices at each M point, giving rise to the sublattice interference mechanism. Dirac cones slightly above the Fermi energy are indicated in red according to their orbital origin.
    }}
\end{figure}

In this Letter, we develop a theory of charge order for
$A$V$_3$Sb$_5$. In order to do so, we obtain an effective two-dimensional tight binding
model which manages to keep the most salient features of the $A$V$_3$Sb$_5$
band structure. We find that, assuming an electronically mediated charge order,
the Fermi pocket at van Hove filling should dominate. On a mean-field level, we compare CBO with $l=1$ against charge density order (CDO) with $l=0$ angular momentum, both for the same ordering wave vectors dictated by the Fermiology. We confirm that the three-fold CBO instability previously found for the kagome Hubbard model at van Hove filling~\cite{PhysRevLett.110.126405,PhysRevB.87.115135} dominates in a large part of parameter space. Beyond the instability level, the three CBO parameters can in principle form condensates which vary in amplitude and relative phase. We discuss the possibility using  a  Ginzburg Landau analysis and find that the CBO instability preferably breaks time-reversal symmetry. All three order parameters appear simultaneously at the instability level, and yet there is a tendency to nematicity through differing phases below $T_{\text{CDW}}$. These results appear to be in line with the measurements from scanning tunneling microscopy.

{\it Multi-orbital effective model.}
Layered kagome metals as the $A$V$_3$Sb$_5$ family are an exciting platform, hosting electronic features like flat bands and nodal lines~\cite{2021PhRvM5c4801O,jiang2020discovery}. The compounds can be treated as effectively two-dimensional~\cite{PhysRevLett.125.247002}, since the electronic features are dominated by the vanadium orbitals crystallizing in a kagome lattice structure. We now discuss the symmetry properties of bands arising from $d$ orbitals in the kagome lattice, neglecting spin-orbit coupling. Located in the wallpaper group $p6mm$ (No. 17), the kagome structure consists of three sublattices that arise from placing atoms on Wyckoff position $3g$ with site-symmetry group $D_{2h}$. The symmetry representations of Bloch states depend on the irreducible representation of the orbitals placed on these Wyckoff positions, as described by the topological quantum chemistry framework~\cite{topquantchem}. Two sets of orbitals contribute:
First, a linear combination of the $d_{\mathrm{xy}}, d_{\mathrm{x}^2-\mathrm{y}^2}, d_{\mathrm{z}^2}$ orbitals forms a Wannier state in the $A_g$ irreducible representation of $D_{2h}$. We focus on the $d_{\mathrm{xy}}$ orbitals as their representative (see Fig.~\ref{fig:lat-bndstruc}a). They induce the well-known kagome band structure with a p-type van Hove singularity~\cite{wu2021nature} at the M point (blue bands in Fig.~\ref{fig:lat-bndstruc}b). 
Second, the $d_{\mathrm{xz/yz}}$ orbitals in the $B_{2g,3g}$ irreducible representations of $D_{2h}$, indicated as two red tones in Fig.~\ref{fig:lat-bndstruc}a, form a set of bands with opposite mirror eigenvalues along the $\Gamma$-M line.
These bands give rise to a mirror-symmetry-protected Dirac cone on the $\Gamma$-M line and additional p- as well as m-type van Hove singularities (red bands in Fig.~\ref{fig:lat-bndstruc}b).
Crossings between the $d_{\mathrm{xy}}$ and the $d_{\mathrm{xz/yz}}$ bands are also protected by mirror symmetry. 

P-type bands discussed above are subject to the sublattice interference~\cite{PhysRevB.86.121105}: at each M point, the Bloch states at van Hove filling have support on only one of the three sublattices (see Fig.~\ref{fig:lat-bndstruc}c). This effect is not an artifact of a simplified effective model, but is also seen in the first principles calculation to very high accuracy (corrected only by a small admixture from $p$ orbtials of the other sublattices). It has a crucial influence on the character of Fermi instabilities~\cite{PhysRevLett.110.126405,PhysRevB.86.121105,PhysRevB.100.085136}. 

In order to unravel the origin of the observed Fermi surface instability, we focus on nesting effects which are most apparent in the $d_{\mathrm{xy}}$ bands (see the nesting wave vectors $\bs{Q}_j$, $j=1,2,3$, indicated in Fig.~\ref{fig:lat-bndstruc}c). We therefore focus on the $d_{\mathrm{xy}}$ orbitals exclusively in the following, while keeping in mind that $d_{\mathrm{xz/yz}}$ bands share their important features -- van Hove singularity and sublattice interference -- and can thus be expected to support the ordering tendencies arising from the $d_{\mathrm{xy}}$ bands (see Sec.I in Ref.~\onlinecite{suppmat}).

 The tight-binding description of the $d_{\mathrm{xy}}$ orbital band structure is $H_0 = \sum_{\bs{k},\alpha,\beta} c_{\bs{k},\alpha}^{\dagger} \mathcal{H}_{\bs{k},\alpha, \beta} c_{\bs{k},\beta}^{}$, where $c_{\bs{k},\alpha}^{\dagger}$ creates
 a Bloch electron with momentum $\bs{k}$ in and support on sublattice $\alpha=1,2,3$. The Bloch Hamiltonian matrix reads 
\begin{equation}
    \mathcal{H}_{\bs{k}} = -2t\begin{pmatrix}
    \mu &  \cos\left(\bs{k}\bs{a}_3\right) & \cos\left(\bs{k}\bs{a}_2\right)\\
     \cos\left(\bs{k}\bs{a}_3\right) & \mu &  \cos\left(\bs{k}\bs{a}_1\right)\\
    \cos\left(\bs{k}\bs{a}_2\right) &  \cos\left(\bs{k}\bs{a}_1\right) & \mu
    \end{pmatrix},
\end{equation}
where we use the lattice vectors connecting sublattices as $\mathbf{a}_{1,3} = \left(1/4, \mp \sqrt{3}/4\right)^{\mathrm{T}}$, $\mathbf{a}_{2} = \left(1/2, 0\right)^{\mathrm{T}}$, $t = 0.3~\mathrm{eV}$ as the overall hopping strength and $\mu$ as the chemical potential. The corresponding bandstructure presented in Fig.~\ref{fig:lat-bndstruc}b shows an enhancement of the density of states around the three $M$-points at the van Hove filling $n = 5/12$. 

\begin{figure}
    \includegraphics[width=\linewidth]{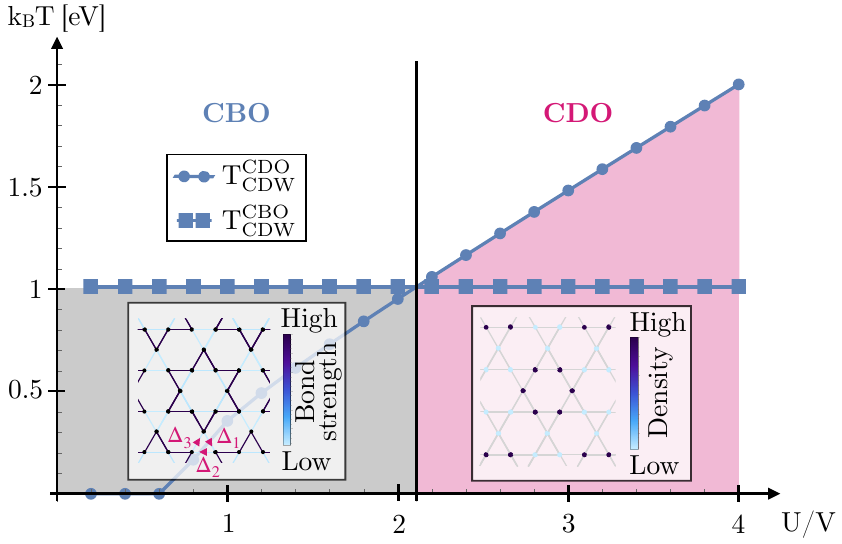}
    \caption{\label{fig:phasediagram} \textsf{\textbf{Interaction strength phase diagram and corresponding electronic order in AV$_3$Sb$_5$.} Mean field critical temperatures of the charge density order (CDO) introduced in Eq.~\eqref{eq:CDO-k} and the charge bond order (CBO) introduced in Eq.~\eqref{eq:CBO-k} as a function of the electronic interaction strengths $U/V$ ($\mu = -35$ meV~\cite{tsirlin2021anisotropic}). For $U \lesssim 2.1 V$ a charge bond order dominates, which forms a complex star of David pattern of strong and weak bonds. The inset highlights the mean field parameters $\Delta_j$, $j=1,2,3$, and their arrangement on the kagome lattice. For systems with interaction scales above $U = 2.1 V$ a charge density order emerges, modulating the real space site occupation. 
    }}
\end{figure}

{\it Charge density wave instability.} 
The emergence of the charge density wave in $A$V$_3$Sb$_5$ has been experimentally observed to break the translational symmetry of the kagome lattice to a $2\times2$ unit cell~\cite{jiang2020discovery,arXiv210309188,tan2021charge,uykur2021optical,arXiv210210959,arXiv210210987,arXiv210110193}, thereby necessitating an instability with $\bs{Q} \neq 0$, specifically $\bs{Q}_{1,3} = \left(\pi, \mp \sqrt{3}\pi\right)^{\mathrm{T}}$, $\bs{Q}_{2} = \left(2 \pi, 0\right)^{\mathrm{T}}$ (see Fig.~\ref{fig:lat-bndstruc}c). Experimental evidence in AV$_3$Sb$_5$ is indicative of an electronically driven charge order~\cite{arXiv2103.09769,zhou2021origin}. This is why we consider a mean-field treatment of the kagome Hubbard model, realized by the Hamiltonian
\begin{equation}
\begin{split}
    H =& H_0 + H_{\text{int}}\\
      =& \mu \sum_{\bs{r},\sigma} c_{\bs{r},\sigma}^{\dagger}c_{\bs{r},\sigma}^{}-t \sum_{\langle \bs{r},\bs{r}' \rangle,\sigma} \left(c_{\bs{r},\sigma}^{\dagger}c_{\bs{r}',\sigma}^{} + \mathrm{h.c.}\right) \\
      &+ U \sum_{\bs{r}} n_{\bs{r},\uparrow}n_{\bs{r},\downarrow} + V \sum_{\langle \bs{r},\bs{r}' \rangle, \sigma, \sigma'} n_{\bs{r},\sigma}n_{\bs{r}',\sigma'},
    \end{split}
\end{equation}
where $c_{\bs{r},\sigma}^\dagger$ creates an electron on site $\bs{r}$ in the kagome lattice with spin $\sigma=\uparrow,\downarrow$ and $n_{\bs{r},\sigma}=c_{\bs{r},\sigma}^\dagger c_{\bs{r},\sigma}^{}$. The on-site Hubbard interaction is parametrized by $U$, while the nearest-neighbor interaction strength is denoted by $V$. Crucially, the hopping always acts between densities on sites belonging to different sublattices. 

In the following, we only consider charge orders, i.e., all fermionic bilinear forms are implicitly summed over their spin degree of freedom. Note that a spin bond order with likewise finite relative angular momentum could in principle be a close competitor~\cite{PhysRevLett.110.126405}, which could evade conventional static measurements of local magnetic moments~\cite{PhysRevB.62.4880} and hence would not contradict e.g. the absence of a local magnetic moment in $\mu$SR~\cite{10.1088/1361-648X/abe8f9}. The current related to the anomalous Hall signal~\cite{Yangeabb6003}, however, manifestly is a charge current and not a spin current. Additionally, recent optical spectroscopy experiments support the bulk nature of the CDW state~\cite{uykur2021optical}, and as such hint at an electronically mediated charge order as its origin. 

\begin{figure*}
    \includegraphics[width=\linewidth]{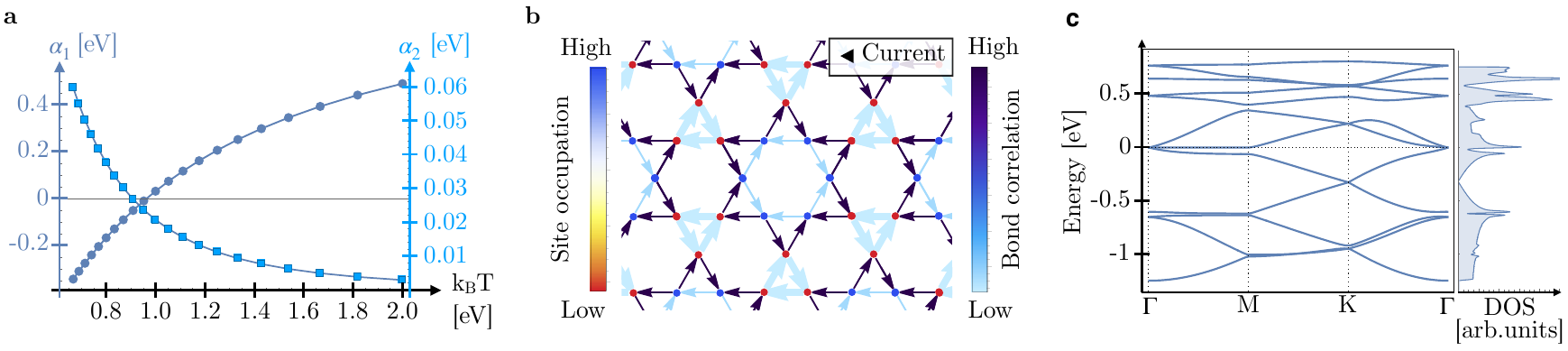}
    \caption{\label{fig:GL-current} \textsf{\textbf{Ginzburg Landau free energy and resulting current pattern.} (\textbf{a}) The second order coefficient $\alpha_1$ as a function of temperature $T$ signals the second order phase transition of the emerging charge bond order by a sign change at $\mathrm{k}_{\mathrm{B}}T_{\mathrm{CDW}} \approx 1.011$~eV, whereas the coefficient $\alpha_2$ is purely positive, requiring an imaginary order parameter $\Delta_j$, breaking time-reversal symmetry ($V=1$~eV, $\mu = -35$~meV). (\textbf{b}) The bond correlation pattern $|\langle \psi_{\mathrm{GS}} | c_{\bs{r}}^{\dagger} c_{\bs{r}'}^{} | \psi_{\mathrm{GS}} \rangle|$ in this setting ($\Delta_j = 0.1\imi$) shows a star of David arrangement of strong and weak bonds, equipped with a non-zero current. The current is varying in size on the lattice, highlighted by the thickness and direction of arrows (thin small, thick large current). Additionally, the order parameter modulates the on-site occupation within and between the hexagons. (\textbf{c}) The resulting bandstructure in the reduced Brillouin zone shows a gap opening around the $M$-points, reducing the density of states (DOS) at the Fermi level ($\Delta_j = 0.1\imi$).
    }}
\end{figure*}

The relative strength of on-site $U$ and nearest neighbor $V$ Hubbard interactions allow for different charge orders to emerge. Inspired by the majority of charge instabilities, a \emph{charge density order} (CDO) is an obvious possibility, described by the order parameter

\begin{equation}
    \mathcal{O}_{\text{CDO}}(\bs{k}) = \sum_\alpha\tilde{\Delta}_\alpha \langle c_{\bs{k},\alpha}^{\dagger} c_{\bs{k} + \bs{Q}_\alpha,\alpha}^{}\rangle,
    \label{eq:CDO-k}
\end{equation}
where $\alpha$ labels the three sublattices. It corresponds to a real space pattern of the form

\begin{equation}
    \mathcal{O}_{\text{CDO}}(\bs{r}) = \sum_\alpha\tilde{\Delta}_\alpha \cos\left(\bs{Q}_\alpha\bs{R}\right) \langle c_{\bs{R},\alpha}^{\dagger} c_{\bs{R},\alpha}^{}\rangle,
    \label{eq:CDO-r}
\end{equation}
where $c_{\bs{R},\alpha}^{\dagger}$ creates an electron in sublattice $\alpha$ of unit cell $\bs{R}$, while $\bs{r}=\bs{R}+\bs{r}_\alpha$ is the actual position of the site ($\bs{r}_1=\bs{0}$, $\bs{r}_2=\bs{a}_3$, $\bs{r}_3=\bs{a}_2$). Such an arrangement is visualized in the right inset of Fig.~\ref{fig:phasediagram}.

Owing to the unique sublattice structure, however, the inhomogeneous distribution of Fermi level density of states reduces nesting effects for a local Hubbard interaction. Consequently, the nearest-neighbor interaction $V$ is promoted, leading to a \emph{charge bond order}, which modulates the kinetic hopping strengths instead of on-site densities. Correspondingly, the order parameter in reciprocal space can be written as~\cite{PhysRevLett.110.126405}

\begin{equation}
    \mathcal{O}_{\text{CBO}}(\bs{k}) = \sum_{\alpha,j,\beta}\Delta_j \sin\left(\frac{\bs{Q}_j\bs{k}}{4\pi}\right) \langle c_{\bs{k},\alpha}^{\dagger} c_{\bs{k} + \bs{Q}_j,\beta}^{}\rangle |\epsilon_{\alpha j \beta}|,
    \label{eq:CBO-k}
\end{equation}
where $\epsilon_{\alpha j \beta}$ is the Levi-Civita tensor and $\alpha,\beta$ run over the three sublattices. The $\bs{k}$-dependence of the order parameter creates a non-trivial relative momentum structure, leading to the formation of an unconventional CDW order. Note that the relative angular momentum $l=1$ of the individual particle hole pairs does not directly carry over to the macroscopic charge order ground state formed by the coherent superposition of these particle hole pairs. Rather, the $l=1$ substructure of the particle hole pair wave function unfolds in the nature of particle-hole excitations above the ground state.
In real space, the emerging order corresponds to an alternating modulation of the hopping strengths connecting the three sublattices, giving rise to the enlarged $2\times 2$ unit cell
\begin{equation}
\begin{split}
    \mathcal{O}_{\text{CBO}}(\bs{r}) = \sum_{\alpha,j,\beta}&\Delta_j \cos\left(\bs{Q}_j\bs{R}\right)|\epsilon_{\alpha j \beta}| \\
    &\left(\langle c_{\bs{R},\alpha}^{\dagger} c_{\bs{R},\beta}^{}\rangle-\langle c_{\bs{R},\alpha}^{\dagger} c_{\bs{R}-2\bs{a}_j,\beta}^{}\rangle\right),
    \end{split}
    \label{eq:CBO-r}
\end{equation}
whose visualization is given in the left inset of Fig.~\ref{fig:phasediagram}. 
We compare the transition temperature of the two orders defined in Eqs.~\eqref{eq:CDO-k} and~\eqref{eq:CBO-k} on the mean-field level. The results are visualized in Fig.~\ref{fig:phasediagram}. Clearly, a CBO is favored for a wide range of interaction strengths, owing to the unique sublattice structure of the kagome lattice~\footnote{Another CBO is discussed in Ref.~\onlinecite{2021arXiv210402725L}, yielding the same critical temperature as obtained here.}. This allows nearest-neighbor interactions $V$ to take the role of mediating the unconventional charge order. By contrast, the CDO is found in the limit of large on-site interaction $U$. Experimentally, a transition temperature $T_{\mathrm{CDW}} \approx 78-103$ K has been observed~\cite{arXiv210309188}. Gauging our scales from this input, it yields $U/2=V\approx250$~meV in our setting. Naturally, for sufficiently large on-site interactions $U$, the charge density order becomes dominant again, as seen on the right side of Fig.~\ref{fig:phasediagram}. Exerting hydrostatic pressure has been found to lead to a suppression of the charge order~\cite{tsirlin2021anisotropic,arXiv210210959}, related to both Fermi surface renormalizations and changes in the nearest-neighbor interaction $V$ (see Sec.II in Ref.~\onlinecite{suppmat}).

{\it Ginzburg Landau formalism.}
To study the interplay of the triplet of order parameters in the CBO phase, we derive the symmetry-constrained expansion of the Ginzburg Landau free energy for $\Delta_1$, $\Delta_2$, $\Delta_3$ as defined in Eq.~\eqref{eq:CBO-r}. The relevant generating symmetries are translation by $2\bs{a}_1$, $    (\Delta_1,\Delta_2,\Delta_3)\to (\Delta_1,- \Delta_2, -\Delta_3)$,
translation by $2\bs{a}_2$, $
    (\Delta_1,\Delta_2,\Delta_3)\to (-\Delta_1, \Delta_2, -\Delta_3)$
six-fold rotation, $
    (\Delta_1,\Delta_2,\Delta_3)\to (\Delta_3^*, \Delta_1^*, \Delta_2^*)$,
mirror with $\bs{a}_2$ normal to the mirror plane, $    (\Delta_1,\Delta_2,\Delta_3)\to (\Delta_3^*, \Delta_2^*, \Delta_1^*)$,
and time-reversal acting as complex conjugation. For a more compact notation, we decompose the complex order parameter into phase $\phi_j$ and absolute value $\psi_j>0$ as $\Delta_j=\psi_je^{i\phi_j}$, for $j=1,2,3$. Up to third order in $\Delta_j$, and neglecting gradient terms, the free energy expansion reads~\cite{PhysRevB.12.1187,van_Wezel_2011}
\begin{equation}
\label{eq: free energy}
    \begin{split}
    F=&
    \alpha_1\sum_j \psi_j^2
    +2\alpha_2\sum_j \psi_j^2 \cos(2\phi_j)
    \\
    &+2\gamma_1\,\psi_1\psi_2\psi_3\,\cos(\phi_1+\phi_2+\phi_3)
    \\
    &+\gamma_2\,\psi_1\psi_2\psi_3\,\bigl[8\cos(\phi_1)\cos(\phi_2)\cos(\phi_3)
    \\
    &\qquad\qquad
    -2\cos(\phi_1+\phi_2+\phi_3)
    \bigr],
    \end{split}
\end{equation}
with the coefficients $\alpha_{1,2}$ and $\gamma_{1,2}$ being temperature-dependent real numbers. We discuss the effect of the second- and third-order terms separately, with the vanishing $\alpha_1\pm 2\alpha_2$ at $T_{\mathrm{CDW}}$ indicating the phase transition to the charge bond ordered state (see Fig.~\ref{fig:GL-current}a).

The second order terms are minimized by $\phi_j\,\mathrm{mod}\, \pi = 0$ for $\alpha_2<0$, and $\phi_j\,\mathrm{mod}\, \pi = \pi/2$ for  $\alpha_2>0$, where the latter case breaks time-reversal symmetry spontaneously and induces orbital currents~\cite{PhysRevB.82.045102,PhysRevLett.117.096402}. Our microscopic calculation indeed yields $\alpha_2>0$, as seen in Fig.~\ref{fig:GL-current}a. The resulting bond correlation $|\langle \psi_{\mathrm{GS}} | c_{\bs{r}}^{\dagger} c_{\bs{r}'}^{} | \psi_{\mathrm{GS}} \rangle|$, with ${\bs{r}}$, $\bs{r}'$ being nearest-neighbor sites and $\psi_{\mathrm{GS}}$ the single Slater determinant groundstate wavefunction, replicates the expected star of David pattern imprinted by the modulated hopping elements. Additionally, orbital currents emerge due to the time-reversal symmetry breaking and, even though we have a bond order, the on-site densities $\langle \psi_{\mathrm{GS}} | c_{\bs{r}}^{\dagger} c_{\bs{r}}^{} | \psi_{\mathrm{GS}} \rangle$ are also modulated (see Fig.~\ref{fig:GL-current}b).

The third order terms in Eq.~\eqref{eq: free energy} mediate interactions between the three order parameters. They demonstrate that the simultaneous nucleation of all three order parameters is energetically favorable. Our microscopic calculation yields $\gamma_1<0$ and $\gamma_2=0$. Minimization of the third order terms alone then implies $\phi_1+\phi_2+\phi_3\,\mathrm{mod}\,2\pi=0$, which forces the $\phi_j$ to deviate from $\pm\pi/2$, while maintaining the time-reversal symmetry breaking. This competition between second and third order terms, together with the fourth order, leads to a transition from an isotropic to an anisotropic charge order upon lowering temperature. Specifically, a difference between the phase factors $\phi_1 \neq \phi_{2,3} \neq \pi/2$ emerges within the charge order phase, while maintaining a complex order parameter with $\psi_1 = \psi_2 =\psi_3$. This results in a \emph{nematic} chiral charge order, where the bond correlations spontaneously break the $C_6$-rotational symmetry (see Sec.III in Ref.~\onlinecite{suppmat}). Such a scenario has been observed in numerous experiments~\cite{jiang2020discovery,li2021rotation,ratcliff2021coherent,wang2021unconventional}.

We conclude from this Ginzburg Landau analysis three points that are in accordance with the experimental findings, in particular in AV$_3$Sb$_5$: (i) the CBO breaks time-reversal symmetry spontaneously, (ii) all three order parameter components nucleate together, and (iii) their competition can result in a nematic CBO. 

In Fig.~\ref{fig:GL-current}c the band structure of the time-reversal symmetry breaking CBO mean-field state is shown. We observe that the order parameter does not open a full gap, but significantly reduces the density of states at the Fermi level inducing two peaks above and below, in accordance with experimental findings~\cite{uykur2021optical,ratcliff2021coherent,cho2021emergence,lou2021chargedensitywaveinduced,hu2021chargeorderassisted,nakayama2021multiple,wang2021distinctive}. 

\vfill
{\it Conclusion and outlook.}
We have developed a minimal tight-binding model capturing the central features relevant for the discussion of charge density wave order in the kagome metals AV$_3$Sb$_5$. Motivated by recent experimental verification, we have investigated the motif of electronically driven charge order through a mean-field treatment of the kagome Hubbard model. Owing to the unique sublattice structure of the Fermi surface instability, a charge bond order with particle hole pairs of non-zero angular momentum emerges. From a Ginzburg Landau expansion of the free energy, we go beyond the instability level and find that the charge order in AV$_3$Sb$_5$ tends to yield orbital currents as a manifestation of time-reversal symmetry breaking, and is naturally prone to the onset of nematicity.  
The recently observed $2\times2\times2$ CDW in (Rb,Cs)V$_3$Sb$_5$~\cite{arXiv2103.09769,arXiv210304760} can likely be explained by extending the CBO order parameter in \eqref{eq:CBO-k} and \eqref{eq:CBO-r} to three dimensions~\cite{PhysRevLett.105.176401}. Intertwining these novel charge orders with the topological features of AV$_3$Sb$_5$ opens a phletora of exciting phenomena, in particular with respect to possibly highly exotic descendant superconducting pairing. 

{\it Acknowledgments.} RT thanks S.~A.~Kivelson for discussions. This work is funded by the Deutsche
Forschungsgemeinschaft (DFG, German Research Foundation) through
Project-ID 258499086 - SFB 1170 and through the W\"urzburg-Dresden
Cluster of Excellence on Complexity and Topology in Quantum Matter - ct.qmat Project-ID 390858490 - EXC 2147. Additionally, this project has received funding from the European Research Council (ERC) under the European Union’s Horizon 2020 research and innovation programm (ERC-StG-Neupert-757867-PARATOP). 

{\it Note added.} Upon completion of this work, we became aware
of Ref.~\onlinecite{feng2021chiral} which investigates
a slightly different charge bond order for AV$_3$Sb$_5$. Repeating our mean-field analysis for this order parameter yields a lower $\mathrm{k}_{\mathrm{B}}T_{\mathrm{CDW}} \approx 423$~meV (for $V = U = 1$~eV), rendering this order energetically less favorable.

\bibliographystyle{apsrev4-1}
\bibliography{bibliography}
\end{document}


\title{Supplementary Material for\\ ``Analysis of Charge Order in the Kagome Metal $A$V$_3$Sb$_5$ ($A=$K,Rb,Cs)"}

\author{M.~Michael~Denner}
\affiliation{Department of Physics, University of Zurich, Winterthurerstrasse 190, 8057 Zurich, Switzerland}
\author{Ronny Thomale}
\affiliation{Institute for Theoretical Physics, University of Würzburg, Am Hubland, D-97074 Würzburg, Germany}
\affiliation{Department of Physics and Quantum Centers in Diamond and Emerging Materials (QuCenDiEM) group, Indian Institute of Technology Madras, Chennai 600036, India}
\author{Titus Neupert}
\affiliation{Department of Physics, University of Zurich, Winterthurerstrasse 190, 8057 Zurich, Switzerland}

\maketitle

\tableofcontents

\section{Nesting instabilities from m-type van Hove singularities}

The single-orbital kagome system of Eq.(1) in the main text possesses two van Hove singularities that provide the necessary density of states for electronic instabilities. However, they differ considerably in their sublattice support around the Fermi surface~\cite{wu2021nature}: The upper van Hove singularity is of p-type, as the eigenstates in the vicinity of the three $M$ points are localized on mutually different sublattices. As such it experiences the sublattice interference mechanism, essential for the formation of the charge bond order. On the other hand, the lower van Hove singularity is of m-type, with the eigenstates equally distributed over mutually different sets of two sublattices for each $M$ point. We investigate the implications of the type of van Hove singularity for the nature of the order parameter by tuning the Fermi level to the m-type van Hove point in our model. 
\begin{figure}[h]
    \includegraphics[width=\linewidth]{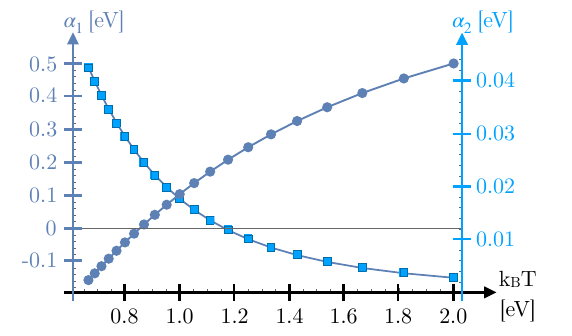}
    \caption{\label{fig:sup2} \textsf{\textbf{Ginzburg Landau free energy second order coefficients at m-type van Hove singularity.} The second order coefficient $\alpha_1$ as a function of temperature $T$ signals the phase transition of the emerging charge bond order by a sign change at $\mathrm{k}_{\mathrm{B}}T_{\mathrm{CDW}} \approx 850$~meV, whereas the coefficient $\alpha_2$ is purely positive, requiring an imaginary order parameter $\Delta_j$, breaking time-reversal symmetry ($V=1$~eV, $\mu = 0.6$). While leading to a lower $T_{\mathrm{CDW}}$, the nature of the order parameter is not altered compared to a p-type Fermi surface nesting.
    }}
\end{figure}
When using the experimentally gauged value of $U/2=V\approx250$~meV, the resulting second order coefficients $\alpha_1$ and $\alpha_2$ in Fig.~\ref{fig:sup2} result in a $T_{\mathrm{CDW}}$ that is a factor of two lower compared with the $T_{\mathrm{CDW}}$ at the filling corresponding to the p-type van Hove singularity. However, $\alpha_2$ still remains positive, favoring a complex order parameter as before. Consequently, while the absence of the sublattice interference mechanism at m-type Fermi surfaces impacts the critical temperature considerably, it does not alter the time-reversal symmetry breaking nature of the order parameter.

The multi-band nature of $A$V$_{3}$Sb$_{5}$ (see Fig.1 of the main text) contains several van Hove singularities, three close to the Fermi level. Two of them are in direct vicinity of the Fermi surface and of p-type, such that the discussion presented in the main text is directly applicable. The slight energetic difference between the two might affect nesting effects slightly, without modifying the nature of the order. The third van Hove singularity sits higher in energy and is of m-type. Therefore, in light of our analysis of isolated m-type van Hove points, we expect the charge ordering not to be affected qualitatively by its presence. 

\section{Pressure dependence of the charge order} 
\begin{figure}[t]
    \includegraphics[width=\linewidth]{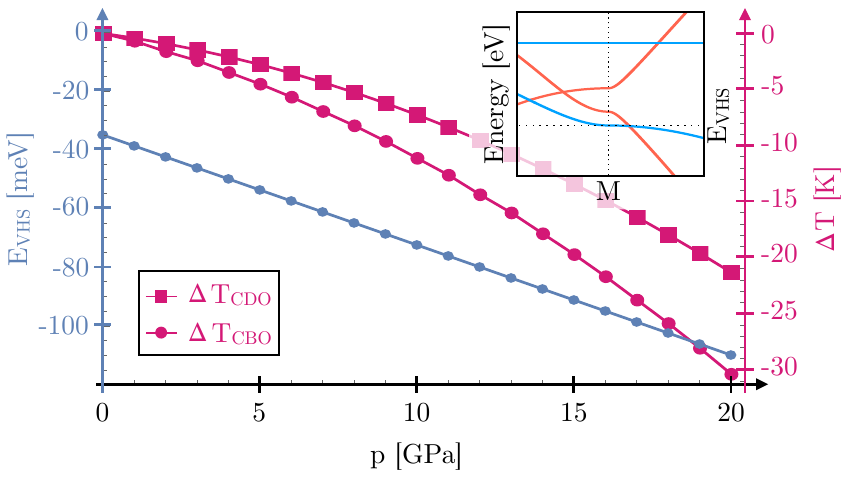}
    \caption{\label{fig:sup1} \textsf{\textbf{Pressure tuneability of van Hove singularity and critical temperature.} The position $E_{\mathrm{VHS}}$ of the p-type van Hove singularity in $A$V$_3$Sb$_5$ (inset) can be tuned by the application of hydrostatic pressure $p$. Increasing $p$ pushes the van Hove point further away from the Fermi level. This alters the Fermi surface and nesting condition, lowering the critical temperature of the COs compared to the ambient pressure case with $E_{\mathrm{VHS}} \approx -35$~meV ($V=1$~eV, $U = 2.1$~eV).
    }}
\end{figure}

The transition to the charge ordered state induces a distortion of the underlying kagome lattice. In the three-dimensional compound $A$V$_3$Sb$_5$, this results in an evasive movement of the encapsulating antimony structure~\cite{tsirlin2021anisotropic}. The application of external pressure hinders this evasive movement, leading to a reduction of the critical temperature $T_{\mathrm{CDW}}$~\cite{du2021}. Viewed from the perspective of the electronic band structure, hydrostatic pressure results in moving the important van Hove singularity further away from the Fermi level~\cite{tsirlin2021anisotropic}. As a result the nesting is reduced, thereby suppressing the on-set of the charge ordered state. In our theoretical setting, we mimic this by changing the term $\mu$ in the Hamiltonian 
\begin{equation}
\begin{split}
    H =& H_0 + H_{\text{int}}\\
      =& \mu \sum_{\bs{r},\sigma} c_{\bs{r},\sigma}^{\dagger}c_{\bs{r},\sigma}^{} -t \sum_{\langle \bs{r},\bs{r}' \rangle,\sigma} \left(c_{\bs{r},\sigma}^{\dagger}c_{\bs{r}',\sigma}^{} + \mathrm{h.c.}\right) \\
      &+ U \sum_{\bs{r}} n_{\bs{r},\uparrow}n_{\bs{r},\downarrow} + V \sum_{\langle \bs{r},\bs{r}' \rangle, \sigma, \sigma'} n_{\bs{r},\sigma}n_{\bs{r}',\sigma'},
    \end{split}
\end{equation}
shifting the energetic position of the van Hove singularity (see Fig.~\ref{fig:sup1}). Using experimentally obtained values from Ref.~\onlinecite{tsirlin2021anisotropic}, we observe a reduction of both the density order driven by the on-site interaction $U$, as well as the bond order driven by the nearest-neighbor terms $V$. The shift of the van Hove singularity therefore serves as an explanation for the expected reduction in $T_{\mathrm{CDW}}$. Nevertheless the phase transition point ($V=1$~eV, $U = 2.1$~eV) shifts slightly, too, such that a reduction of $V$ could lead to a suppression of the unconventional charge order discussed here. However, the suppression of bond orders opens the possibility of other charge density waves emerging, as indicated by some recent experiments~\cite{chen2021double} as well as theoretical calculations~\cite{zhang2021firstprinciples}.

\section{Nematicity of the order parameter}

The expansion of the Ginzburg Landau free energy allows not only to investigate the critical temperature, but also the precise form of the order parameter. As explained in the main text, the second order term determines the on-set temperature as well as the tendency to break time-reversal symmetry. The third order terms in Eq.(7) of the main text mediate interactions between the three order parameters. The resulting competition between second and third order terms is settled by the fourth order, determining stable solutions. The extension of the expansion from the main text to fourth order reads
\begin{equation}
\label{eq:free energy full}
    \begin{split}
    F=&
    \alpha_1\sum_j \psi_j^2
    +2\alpha_2\sum_j \psi_j^2 \cos(2\phi_j)
    \\
    &+2\gamma_1\,\psi_1\psi_2\psi_3\,\cos(\phi_1+\phi_2+\phi_3)
    \\
    &+\gamma_2\,\psi_1\psi_2\psi_3\,\bigl[8\cos(\phi_1)\cos(\phi_2)\cos(\phi_3)
    \\
    &\qquad\qquad
    -2\cos(\phi_1+\phi_2+\phi_3)
    \bigr]\\
    &+\delta_1\sum_j \psi_j^4 + 2 \delta_2 \sum_j \psi_j^4 \cos(4\phi_j) \\
    &+ \delta_3 \sum_{i,j>i} \psi_i^2\psi_j^2+2\delta_4  \sum_j \psi_j^4 \cos(2\phi_j)\\
    &+2\delta_5\sum_{i,j>i} \psi_i^2\psi_j^2\cos(2(\phi_i+\phi_j))\\
    &+2\delta_6\sum_{i,j>i} \psi_i^2\psi_j^2\cos(2(\phi_i-\phi_j))\\
    &+2\delta_7  \sum_{i,j\neq i} \psi_i^2 \psi_j^2 \cos(2\phi_j).
    \end{split}
\end{equation}

\begin{figure*}
    \includegraphics[width=\linewidth]{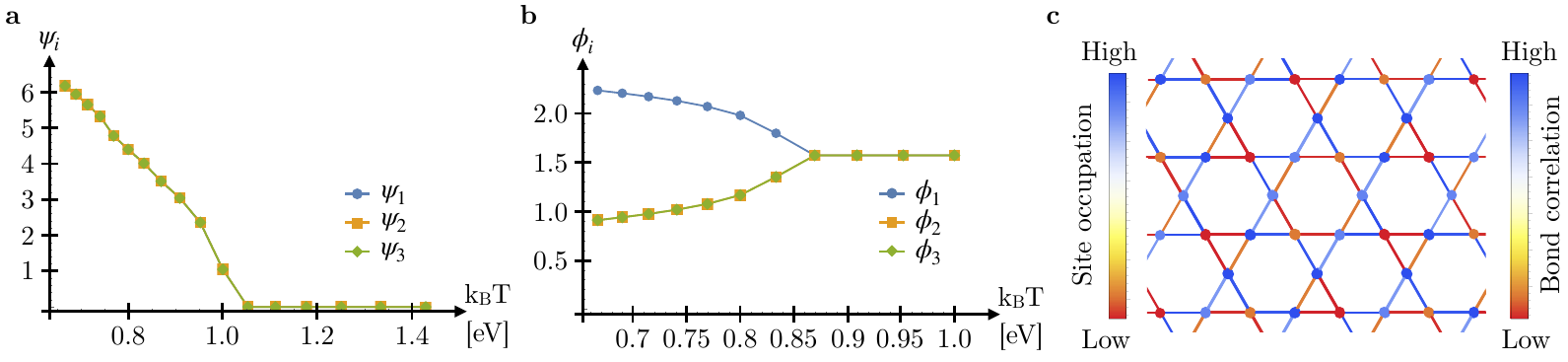}
    \caption{\label{fig:sup3} \textsf{\textbf{Nematicity of the order parameter.} (\textbf{a}) The solution $\Delta_j = |\psi_j| e^{i\phi_j}$ minimizing the free energy~\eqref{eq:free energy full} shows a non-zero order parameter below the critical temperature $\mathrm{k}_{\mathrm{B}}T_{\mathrm{CDW}} \approx 1.011$~eV ($V=1$~eV, $\mu = -35$~meV). (\textbf{b}) Due to the positive coefficient $\alpha_2$, the order parameter carries a non-trivial phase $\phi_j$, which differs between the components $\phi_1 \neq \phi_{2,3}$ below the transition point. (\textbf{c}) The resulting bond correlation and on-site density within the CO phase shows an anisotropy between different directions, explaining the experimentally observed breaking of $C_6$ rotational symmetry ($\mathrm{k}_{\mathrm{B}}T_{\mathrm{CDW}} \approx 0.6$~eV, $|\psi_j| = 6.2$, $\phi_1 = 2.4$, $\phi_{2,3} = 0.7$). The current pattern is not displayed for simplicity.
    }}
\end{figure*}

Numerical optimization of the prefactors $\alpha_i, \gamma_j, \delta_k$ reveals a complex order parameter ($\phi_j = \pi/2$) right below the critical temperature, which is isotropic, carrying the same phase and magnitude for each component (see Fig.~\ref{fig:sup3}a and b). Upon lowering the temperature further, a tendency to nematicity evolves below $\mathrm{k}_{\mathrm{B}}T \approx 0.86$~eV, i.e., the order parameters differ in their phase $\phi_1 \neq \phi_{2,3}$ (see Fig.~\ref{fig:sup3}a and b) while maintaining the same magnitude. As illustrated in Fig.~\ref{fig:sup3}c, this leads to an anisotropy of both on-site densities and bond correlation $|\langle \psi_{\mathrm{GS}} | c_{\bs{r}}^{\dagger} c_{\bs{r}'}^{} | \psi_{\mathrm{GS}} \rangle|$, breaking the inherent $C_6$-rotational symmetry spontaneously. Consequently, we observe a transition between an isotropic to an anisotropic CO upon cooling, while retaining the time-reversal symmetry breaking nature. Such a scenario can be even enriched when considering the three dimensionality of the material, introducing additional complexity along the $c-$axis. These results are in accordance with experimental observations in $A$V$_3$Sb$_5$, showing a strong anisotropy~\cite{jiang2020discovery,li2021rotation,ratcliff2021coherent,wang2021unconventional}.

\newpage
\bibliographystyle{apsrev4-1}
\bibliography{supp-bib}